# Social bot detection in the age of ChatGPT: Challenges and opportunities

by Emilio Ferrara

## Abstract

We present a comprehensive overview of the challenges and opportunities in social bot detection in the context of the rise of sophisticated AI-based chatbots. By examining the state of the art in social bot detection techniques and the more salient real-world application to date, we identify gaps and emerging trends in the field, with a focus on addressing the unique challenges posed by AI-generated conversations and behaviors. We suggest potentially promising opportunities and research directions in social bot detection, including (i) the use of generative agents for synthetic data generation, testing and evaluation; (ii) the need for multimodal and cross-platform detection based on network and behavioral signatures of coordination and influence; (iii) the opportunity to extend bot detection to non-English and low-resource language settings; and, (iv) the room for development of collaborative, federated learning detection models that can help facilitate cooperation between different organizations and platforms while preserving user privacy.

**Contents**

## 1. Introduction

The surge in artificial intelligence (AI) and natural language processing (NLP) technologies has given rise to advanced social bots, presenting new challenges in online communication and cybersecurity. These bots mimic human behavior and interact across platforms, often with possible malicious intent like spreading disinformation or manipulating public sentiment. The advent of AI-generated chatbots like ChatGPT has amplified these issues, making social bot detection and mitigation an imperative task.

This paper offers an in-depth review of social bot detection, encapsulating the challenges, prospects, and emerging trends in this area. We trace the evolution of bot detection techniques from initial heuristic-based methods to the employment of machine learning, NLP, and deep learning approaches. The influence of advanced language models like ChatGPT on detection techniques is also discussed. We then address the complexities in social bot detection such as the sophistication of AI-generated content, adversarial maneuvers, scalability, real-time detection, and ethical and privacy considerations. Concurrently, we highlight future directions like the use of transfer learning, unsupervised learning, multimodal approaches, collaborative learning, and explainable AI in detection methods.

Furthermore, we introduce real-world cases demonstrating the application of bot detection in areas such as election interference, disinformation campaigns, online conspiracy theories, and financial fraud. We also discuss the potential role of generative agents and synthetic data in advancing this field. In conclusion, we propose best practices for social bot detection. This involves integrating diverse detection techniques, data sources and modalities, periodically refining models, maintaining privacy in federated contexts, and harnessing generative agents and synthetic data.

### *1.1. Background on social bots and their role in society*

Social bots, or automated software mimicking human behavior on social media platforms, operate via algorithms ranging from simple scripts to complex AIs, engaging in tasks or human-like conversations online (Ferrara, *et al.*, 2016). With the rise of social media, their use, both benign and malicious, has grown.

Benign social bots help automate content sharing, news aggregation, customer service, and support, thus aiding businesses in managing their social media presence and enhancing user engagement (Ferrara, 2020; Mønsted, *et al.*, 2017; Brandtzaeg and Følstad, 2017).

Conversely, malicious social bots, which garner more attention due to their potential harm to online communities and ability to manipulate public sentiment, have been involved in disinformation spread, online conspiracy proliferation, and political interference (Bessi and Ferrara, 2016; Shao, *et al.*, 2018; Ferrara, 2020; Wang, *et al.*, 2023). As such, their detection and mitigation have become crucial.

The sophistication of social bots has progressed from early rule-based, keyword-driven interactions to the incorporation of machine learning and natural language processing, allowing for more realistic human-like exchanges (Chu, *et al.*, 2010; Chang and Ferrara, 2022). For example, some bots employ Markov chain-based models to generate text, mimicking human-produced content (Hwang, *et al.*, 2012; Pozzana and Ferrara, 2020).

Deep learning and neural network advancements have further enhanced bots' language abilities (Radziwill and Benton, 2017). AI-powered bots, such as OpenAI's ChatGPT, can grasp complex human communication patterns, generating increasingly indistinguishable responses from actual human conversations, thus elevating social bot capabilities.

### *1.2. The rise of AI-generated chatbots like ChatGPT*

The introduction of advanced AI technologies has facilitated the creation of sophisticated chatbots like OpenAI's ChatGPT. Based on GPT-4 architecture, an expansion of the original GPT (Radford, *et al.*, 2018), ChatGPT generates coherent and context aware text that closely resembles human conversation. This architecture employs a deep learning model known as the Transformer (Vaswani, *et al.*, 2017), highly effective for tasks involving natural language understanding and generation.

Owing to its large scale and extensive training data, ChatGPT is trained on hundreds of billions of words and fine-tuned using reinforcement learning from human feedback (RLFH) (OpenAI, 2021). The combination of unsupervised pre-training and reinforcement learning-based fine-tuning allows ChatGPT to generate contextually appropriate and human-aligned responses.

The progress in natural language generation exhibited by ChatGPT and other large-scale language models such as BERT (Devlin, *et al.*, 2018) and XLNet (Yang, *et al.*, 2019) has considerably enhanced social bots' abilities. This enables more sophisticated and nuanced user interactions, opening new AI application opportunities but also intensifying misuse potential, thus complicating social bot detection.

### *1.3. The importance of social bot detection*

Recognizing and differentiating between human users and social bots is pivotal to preserving the integrity of online communities and social media platforms. Effective social bot detection enables the reduction of malicious bot impacts, such as disinformation spread, public opinion manipulation, and harmful content amplification.

Malicious social bots exploit social media platforms to spread disinformation and sway public discourse (Ferrara, 2017; Shao, *et al.*, 2018). In political scenarios like elections, these bots disseminate false information to manipulate public sentiment (Bessi and Ferrara, 2016; Howard and Kollanyi, 2016). Reliable social bot detection can help uncover and counteract these disinformation campaigns, promoting a transparent and democratic online space.

Online conspiracy theories are also perpetuated by social bots (Samory and Mitra, 2018; Muric, *et al.*, 2021). These bots can amplify damaging content and carry out targeted campaigns. Employing robust bot detection mechanisms can help safeguard users from such malicious activities, ensuring a secure and inclusive online environment.

Social media platforms' policies prohibit automated accounts from malicious activities or manipulating platform metrics such as likes, shares, and followers (Ferrara, 2022). Enforcing these policies effectively requires detecting and removing social bots, thus enhancing user experience authenticity. Additionally, social bots may violate user privacy by unauthorized data collection or surveillance (Gorwa and Guilbeault, 2020), underscoring the importance of robust social bot detection.

While techniques for social bot detection vary from simple rule-based methods and feature engineering approaches (Chu, *et al.*, 2010; Subrahmanian, *et al.*, 2016) to advanced machine learning and deep learning techniques (Ferrara, *et al.*, 2016; Z. Yang, *et al.*, 2019), the complexity of AI-driven chatbots like ChatGPT introduces new detection challenges, underscoring the importance of ongoing research in this field (Abdullah, *et al.*, 2022; Hajli, *et al.*, 2022).

### *1.4. Scope and objectives of the paper*

This viewpoint paper aims to provide a detailed analysis of social bot detection in the context of the rising sophistication of AI-generated chatbots like ChatGPT. It will focus on the challenges and opportunities that this new chatbot generation presents for bot detection, emphasizing the necessity for innovative solutions and stakeholder collaboration.

- ***Challenges in bot detection***: The intricate nature of AI-powered chatbots like ChatGPT presents significant obstacles to traditional bot detection techniques, such as rule-based systems and feature engineering approaches (Cresci, *et al.*, 2020). Researchers are turning towards advanced machine learning and deep learning techniques to tackle this issue (Yang, *et al.*, 2019; Varol, *et al.*, 2017). This paper will delve into existing detection limitations and discuss possible solutions.
- ***Opportunities for novel approaches***: The fast-paced AI advancements open up avenues for innovative bot detection techniques. Techniques like transfer learning (Pan and Yang, 2010), graph-based analysis, and active learning (Settles, 2010) could potentially enhance detection performance. These emerging trends and their potential contribution to social bot detection will be explored in this paper.
- ***Best practices and future directions***: The paper will highlight best practices for social bot detection, including blending different detection techniques, ongoing model updates, and addressing ethical considerations in detection methods. It will also explore potential future trends in social bot detection, such as the adoption of explainable AI techniques (Adadi and Berrada, 2018) and the integration of human-in-the-loop approaches (Holzinger, 2016).

## 2. The evolution of social bot detection techniques

### *2.1. Early detection methods based on simple heuristics*

In the early stages of social bot detection, researchers primarily relied on simple heuristics and rule-based systems to identify bots (Ferrara, *et al.*, 2016; Cresci, 2020; Orabi, *et al.*, 2020). These techniques focused on exploiting easily observable patterns and features that were indicative of bot-like behavior. Some of the commonly used heuristics included:

- ***Account activity***: Bots typically exhibit high-frequency messaging and account activity compared to human users. By analyzing the number of posts, retweets, or messages per unit of time, researchers were able to identify potential bots (Benevenuto, *et al.*, 2010).
- ***Account metadata***: Certain metadata associated with user profiles, such as account creation date, number of followers, and following-to-follower ratio, can be indicative of bot behavior. For instance, bots often have a disproportionately high following-to-follower ratio and a short account lifespan (Chu, *et al.*, 2010; Wang, *et al.*, 2012).
- ***Content-based features***: Bots often generate content that is repetitive, contains specific keywords, or derived from a limited set of sources. By analyzing the diversity of content, the presence of certain keywords, and the distribution of sources, researchers could identify potential bots (Ratkiewicz, *et al.*, 2011; Lee, *et al.*, 2011).
- ***Network-based features***: Bots often exhibit distinct network patterns, such as forming tightly connected groups or having few reciprocal relationships. By analyzing the structure of the follower and friend networks, researchers could detect potential bot accounts (Stringhini, *et al.*, 2010).

While these early detection methods provided a foundation for identifying social bots, they were limited in their ability to adapt to the evolving sophistication of bot behavior. More advanced bots could easily circumvent these heuristic-based methods by mimicking human behavior, adjusting their activity patterns, or generating diverse content (Chang and Ferrara, 2022).

Furthermore, the reliance on manually crafted rules and features made these methods susceptible to false positives and negatives, as genuine users might exhibit bot-like behavior or *vice versa*. This limitation led to the development of more advanced techniques that incorporated machine learning and natural language processing to improve detection accuracy and adaptability (Subrahmanian, *et al.*, 2016; Ferrara, *et al.*, 2016).

### *2.2. The incorporation of machine learning and natural language processing*

As social bots became more sophisticated, detection methods evolved to include machine learning and natural language processing (NLP) techniques. These innovative approaches aimed to identify bots by analysing nuanced patterns in their behavior, language use, and network features, thus improving accuracy and adaptability compared to traditional heuristic-based methods.

Machine learning algorithms, such as decision trees, support vector machines, logistic regression, and random forests, were used to categorize accounts as bots or humans based on distinct features derived from their online behavior (Subrahmanian, *et al.*, 2016; Ferrara, *et al.*, 2016). By training models on datasets of verified bot and human accounts, researchers designed classifiers that could generalize to unfamiliar instances and adjust to the evolving behavior of social bots. The effectiveness of machine learning-based detection largely depended on the quality of classification features, which include account metadata, content-based features, and network-based attributes. Linguistic features of bot-generated content, such as sentiment scores, lexical diversity, and topic distributions, were also used (Davis, *et al.*, 2016; Kudugunta and Ferrara, 2018).

In situations where labeled data were limited or costly to procure, researchers turned to unsupervised and semi-supervised learning techniques for bot detection. Techniques like clustering, outlier detection, and label propagation sought to identify bots by analyzing inherent data structures and patterns without the need for explicit labels (Chavoshi, *et al.*, 2016; Cao, *et al.*, 2014). These methods can discover novel bot behavior and patterns overlooked by supervised learning techniques.

Moreover, researchers applied NLP techniques to analyze the linguistic features of bot-generated text. Sentiment analysis, topic modeling, and syntactic parsing enabled the extraction of higher-level linguistic patterns and structures (Beskow and Carley, 2018; Addawood, *et al.*, 2019). By integrating NLP features into machine learning models, researchers could more effectively distinguish between bot-generated and human-generated content.

### *2.3. Deep learning and neural network-based approaches*

The emergence of deep learning and neural network technologies has revolutionized the field of social bot detection. These advanced techniques can learn intricate patterns and features from raw data, eliminating the need for manual feature engineering. They have been used to examine textual content and metadata associated with user profiles and activities, offering enhanced detection capabilities over traditional machine learning and NLP methods.

*Convolutional neural networks* (CNNs) have been deployed to examine textual content and extract sophisticated features from word or character sequences. Composed of multiple layers of convolutions and pooling operations, these networks can discern local patterns and hierarchical structures in the data (Zhang, *et al.*, 2015). Researchers have harnessed CNNs to detect social bots by studying the textual patterns and features in their generated content (Min, *et al.*, 2017; Cresci, *et al.*, 2017). *Recurrent neural networks* (RNNs) and *Long short-term memory* (LSTM) networks (Hochreiter and Schmidhuber, 1997; Chung, *et al.*, 2014) have been used to analyze temporal dependencies and sequences in social bot-generated content. These networks excel at learning long-range dependencies and capturing contextual data, making them ideal for analyzing text sequences, user activities, or time series data. They've been used to identify social bots based on the patterns and structures in their generated content and temporal activity (Kudugunta and Ferrara, 2018). *Graph neural networks* (GNNs) have been used to model network structure and interactions between users on online platforms. Designed to detect complex relational patterns and dependencies in graph-structured data, these networks have been employed for detecting social bots based on network properties and relational patterns (Guo, *et al.*, 2021).

Finally, researchers have utilized transfer learning and pre-trained language models, such as BERT, GPT, and RoBERTa (Devlin, *et al.*, 2018; Radford, *et al.*, 2018; Liu, *et al.*, 2019), for social bot detection. Pre-trained on extensive text corpora and fine-tuned for specific tasks, these models can extract rich semantic and syntactic information from text. Leveraging these pre-trained models enhances the detection capabilities and aids in distinguishing between human and bot-generated content (Heidari and Jones, 2020; Guo, *et al.*, 2021). Transfer learning has shown promise in adapting existing bot detection models to low-resource language settings, improving detection performance even under stringent data constraints (Haider, *et al.*, 2023).

### *2.4. The impact of large language models on detection techniques*

The rise of AI-generated chatbots like ChatGPT has presented new challenges for social bot detection. These sophisticated models can produce human-like text with remarkable coherence, fluency, and context-awareness (Radford, *et al.*, 2019). Consequently, traditional detection methods that depend on linguistic patterns or content-based features might struggle to differentiate AI-generated content from human-generated content (Gehrmann, *et al.*, 2019; Zellers, *et al.*, 2019; Grimme, *et al.*, 2022).

Researchers have begun to develop innovative detection methods and techniques to address the challenges presented by ChatGPT and similar AI-generated chatbots. One strategy for detecting AI-generated content is to look for adversarial examples or artifacts specific to the model's training data or architecture. By studying the anomalies and biases in the generated text, researchers may identify patterns or features indicative of AI-generated content (Ippolito, *et al.*, 2020; Solaiman, *et al.*, 2019). Moreover, fine-tuning pre-trained language models on datasets containing both AI-generated and human-generated content could improve these models' detection abilities (Guo, *et al.*, 2021).

As AI-generated text becomes increasingly sophisticated, researchers are drawing parallels between text-based social bots and deepfake technologies that generate realistic, manipulated images or videos. Techniques developed for deepfake detection, like forensic features, ensemble learning, and adversarial training, might be adapted to detect AI-generated text (Afchar, *et al.*, 2018; Li, *et al.*, 2020). Researchers are exploring explainable AI (XAI) techniques and feature attribution methods, such as LIME, SHAP, and Integrated Gradients (Ribeiro, *et al.*, 2016; Lundberg and Lee, 2017; Sundararajan, *et al.*, 2017), to better understand the decision-making process of deep learning models and enhance their detection capabilities. By revealing the factors contributing to the model's predictions, these methods can help identify novel features and patterns indicative of AI-generated content, thereby informing the design of more effective detection techniques.

As AI-generated chatbots continue to advance, they might be used in multimodal contexts, such as generating text and images simultaneously or engaging in interactive conversations. Researchers have started investigating multimodal detection techniques capable of analyzing and integrating information from various modalities to enhance detection accuracy (Cao, *et al.*, 2023).

## 3. Novel challenges in social bot detection

In this section, we discuss the novel challenges that have emerged in social bot detection because of advancements in artificial intelligence and natural language processing technologies. These challenges include the increasingly sophisticated AI-generated content, adversarial attacks, and evasion tactics employed by malicious actors, the need for scalable and real-time detection methods, and the ethical considerations and privacy concerns that must be addressed. By examining these challenges, we aim to provide a deeper understanding of the complexities involved in detecting and mitigating the impact of social bots in the digital age.

| Challenge | Description | Research Opportunities |
|---|---|---|
| **Increasingly sophisticated AI-generated content** | As AI-generated content becomes more realistic and human-like, it becomes more difficult to distinguish social bots from human users. | Develop advanced models capable of identifying subtle differences between AI-generated content and genuine human interactions. |
| **Adversarial attacks and evasion tactics** | Social bots can employ advanced tactics to evade detection, such as mimicking human behavior or altering their patterns. | Investigate adaptive and robust detection techniques that can counteract evolving adversarial strategies and respond to new tactics. |
| **Scalability and real-time detection** | Effective social bot detection needs to be scalable to handle large volumes of data and operate in real-time to mitigate potential damage. | Explore efficient algorithms and distributed computing solutions for large-scale, real-time social bot detection. |
| **Ethical considerations and privacy concerns** | Social bot detection techniques must balance the need for accuracy with preserving user privacy and avoiding false positives that could harm innocent users. | Research privacy-preserving detection methods and ethical guidelines to ensure responsible and fair social bot detection practices. |

### *3.1. Increasingly sophisticated AI-generated content*

The emergence of advanced AI-generated chatbots, such as ChatGPT and its successors, can lead to a significant increase in the sophistication of social bot-generated content. Traditional detection techniques that rely on content-based features or shallow learning approaches may struggle to separate human-generated from AI-generated content.

Several recent studies have highlighted the challenges posed by sophisticated AI-generated content in the context of social bot detection:

- The GLTR study (Gehrmann, *et al.*, 2019) found that even language models as simple as GPT-2 can generate text that is difficult to distinguish from human-written text, even for expert annotators. Presumably, future iterations of GPT-based language models, including ChatGPT and GPT-4 have narrowed the gap between human-generated and AI-generated content, posing challenges for detection techniques based on linguistic patterns or content-based features.

- The GROVER study (Zellers, *et al.*, 2019) demonstrated that AI-generated text can be highly effective at evading detection by humans and automated classifiers. The researchers found that pre-trained language models, when fine-tuned for the task of generating and detecting fake accounts, were both more effective at generating misleading content and more robust to adversarial attacks.

- The detection of AI-generated content in multimodal contexts, such as image captions or video descriptions, presents additional challenges due to the need to integrate and analyze information from multiple modalities.

To address these challenges, researchers have begun to explore novel detection techniques and methodologies that can effectively differentiate between human-generated and AI-generated content. Identifying adversarial examples or artifacts specific to the model's training data or architecture can help detect AI-generated content (Solaiman, *et al.*, 2019; Wolff and Wolff, 2020; Ippolito, *et al.*, 2020). Leveraging the knowledge contained in pre-trained language models, such as BERT, GPT, and RoBERTa, can enhance the detection capabilities of classifiers by capturing rich semantic and syntactic information present in text (Cresci, *et al.*, 2017; Heidari and Jones, 2020; Guo, *et al.*, 2021). Employing explainable AI techniques and feature attribution methods can provide insights into the factors that contribute to the model's predictions, helping to identify novel features and patterns indicative of AI-generated content (Ribeiro, *et al.*, 2016; Lundberg and Lee, 2017; Sundararajan, *et al.*, 2017).

### *3.2. Adversarial attacks and evasion tactics*

As social bot detection techniques evolve, so do the tactics employed by malicious actors to evade detection. Social bots are becoming more adept at avoiding detection by employing adversarial attacks and evasion strategies, posing significant challenges for researchers and practitioners in the field. Some of the key tactics and challenges in this area include:

- ***Mimicking human behavior***: Social bots can increasingly adapt their behavior to resemble human users, making them harder to detect using behavioral patterns or network-based features (Chang and Ferrara, 2022). Techniques such as sentiment analysis, topic modeling, and user profiling have been leveraged to better distinguish between genuine human behavior and bot behavior (Chavoshi, *et al.*, 2016; Cresci, *et al.*, 2017).

- ***Dynamic content generation***: Social bots can employ more diverse and complex content generation strategies, exploiting the capabilities of advanced AI-generated chatbots like ChatGPT to produce context-aware, coherent, and human-like text (Zellers, *et al.*, 2019). This necessitates the development of novel content-based detection techniques that can identify AI-generated text and differentiate it from human-generated content.

- ***Adversarial machine learning***: Malicious actors can exploit the vulnerabilities of machine learning-based detection techniques by crafting adversarial examples that are designed to deceive classifiers (Biggio, *et al.*, 2013). Researchers need to develop more robust and resilient detection techniques that can withstand adversarial attacks, employing methods such as adversarial training, ensemble learning, and data augmentation (Szegedy, *et al.*, 2013; Goodfellow, *et al.*, 2014; Tramèr, *et al.*, 2017).

- ***Camouflaging and obfuscation***: Social bots may employ various camouflaging and obfuscation tactics, such as changing their posting patterns, altering their network structure, or using different communication channels to avoid detection (Ferrara, *et al.*, 2016; Grimme, *et al.*, 2018). Researchers must continuously monitor and adapt their detection methods to address these evolving threats, incorporating new features and updating their models as necessary.

### *3.3. Scalability and real-time detection*

Detecting social bots in large-scale and dynamic online environments presents significant challenges in terms of scalability and real-time detection. Popular social media platforms generate vast amounts of data, requiring detection techniques that can process and analyze this information quickly and efficiently. Furthermore, the rapid dissemination of information on social networks necessitates real-time or near-real-time detection to mitigate the impact of malicious social bots before their activity can cause significant harm. This is particularly true for harmful content, since negative, inflammatory, and false rumors spread faster (Ferrara and Yang, 2015; Stella, *et al.*, 2018; Vosoughi, *et al.*, 2018).

Many advanced detection techniques, particularly those based on deep learning and neural networks, require significant computational resources and can be time-consuming to train and execute. For example, training large-scale transformer models, such as BERT or GPT, involves substantial computational overhead, making it difficult to deploy these models for real-time social bot detection (Devlin, *et al.*, 2018; Radford, *et al.*, 2019).

To address these challenges, researchers have explored various approaches to improve the scalability and efficiency of social bot detection techniques:

- ***Model compression and distillation***: Techniques such as model pruning, quantization, and

knowledge distillation can be employed to reduce the size and computational complexity of deep learning models, enabling more efficient deployment in real-time detection scenarios (Buciluǎ, *et al.*, 2006). These methods can help maintain the accuracy of the model while reducing the computational overhead associated with training and inference.

- ***Incremental learning and online algorithms***: Incremental learning techniques and online algorithms can adapt to new data as it becomes available, allowing for more efficient detection in dynamic environments (JafariAsbagh, *et al.*, 2014). These approaches can update the model incrementally, reducing the need for costly retraining and enabling real-time or near-real-time detection of social bots.

- ***Parallel and distributed processing***: Parallel and distributed processing techniques can be used to harness the computational power of multiple processors or machines, enabling the efficient processing and analysis of large-scale social media data (Gao, *et al.*, 2015). These approaches can help scale social bot detection techniques to handle the massive data volumes generated by popular social media platforms.

- ***Stream-based processing and data reduction***: Stream-based processing techniques can be employed to analyze data in real-time as it is generated, allowing for more efficient detection of social bots in dynamic online environments (Morstatter, *et al.*, 2013; JafariAsbagh, *et al.*, 2014; Gao, *et al.*, 2015). Data reduction techniques, such as sampling, sketching, and aggregation, can also be used to minimize the amount of data that needs to be processed and stored, improving the efficiency of detection techniques.

### *3.4. Ethical considerations and privacy concerns*

Social bot detection techniques navigate a complex ethical and privacy landscape. It's crucial to ensure that detecting and mitigating malicious bots does not violate genuine users' rights or compromise their personal information. Therefore, ethical and privacy-related challenges must be tackled during the development and deployment of social bot detection techniques.

A primary concern in social bot detection is the risk of false positives, where legitimate users are misidentified as bots. This could result in unjust consequences for innocent users, like account suspension or content removal, potentially infringing on their rights to free speech and access to information (Pierri, *et al.*, 2022). To minimize this risk, researchers must focus on developing detection techniques that are exceptionally accurate and extensively validated on diverse and representative datasets. Moreover, transparency and explainability in the decision-making process are paramount to ensure fair and just detection and mitigation of social bots (Ribeiro, *et al.*, 2016).

Detecting social bots often involves analyzing user-generated content, metadata, and behavioral patterns, which raises significant privacy and data protection concerns. To protect user privacy, detection techniques should comply with data protection regulations like the General Data Protection Regulation (GDPR) in the European Union and should minimize the collection, storage, and processing of sensitive personal information. The use of privacy-preserving techniques, such as differential privacy and federated learning, can help protect user data during the analysis process (Dwork, 2006; Ezzeldin, *et al.*, 2021) but has limitations (Fung, *et al.*, 2020).

The rapid progress in AI-generated content presents both opportunities and challenges for social bot detection. While these models can enhance detection techniques' accuracy and robustness, they can also be misused by malicious actors to create more advanced and elusive social bots. Consequently, researchers and practitioners need to ensure that the development and deployment of AI-generated content for social bot detection are ethically and responsibly carried out, with proper safeguards to prevent misuse.

Transparency, accountability, and public trust must be at the forefront of social bot detection techniques' development and deployment. This can be accomplished by openly sharing information about the detection methods, underlying algorithms, and data sources, while validating the techniques on diverse and representative datasets (Holzinger, *et al.*, 2017; Mitchell, *et al.*, 2019). Involving stakeholders, such as platform users and affected communities, in the design, evaluation, and deployment of social bot detection techniques can help ensure their perspectives and concerns are considered, thus fostering trust and acceptance.

## 4. Opportunities and emerging trends

| Opportunity | Description |
|---|---|
| **Leveraging transfer learning and unsupervised learning** | Utilize pre-trained models and unsupervised techniques to improve social bot detection, especially in low-resource languages and domains. |
| **Multimodal approaches to detection** | Combine diverse data modalities (text, images, audio, video) to enhance detection robustness and accuracy by capturing additional signals. |
| **Collaborative and federated learning for detection** | Employ decentralized learning techniques to enhance detection capabilities while preserving user privacy and data security. |
| **Explainable AI and interpretability in detection techniques** | Develop transparent and interpretable detection models to facilitate trust, understanding, and further improvement of detection systems. |
| **Combining multiple detection techniques** | Integrate various detection methods (e.g., rule-based, machine learning, deep learning) to leverage their strengths and mitigate individual weaknesses, thereby enhancing overall detection performance. |
| **Constantly updating and fine-tuning models** | Regularly update and fine-tune detection models with the latest data and techniques to adapt to the evolving landscape of social bots and maintain high detection performance. |
| **Generative agents as source of synthetic data** | Leverage generative agents that simulate believable human behavior to create diverse and rich synthetic data, which can be used to train and improve social bot detection models. |

In this section, we explore the opportunities and emerging trends that are shaping the future of social bot detection. These developments present innovative ways to counter the challenges posed by increasingly sophisticated AI-generated content and adversarial tactics employed by social bots. We will discuss the potential of transfer learning and unsupervised learning, multimodal approaches to detection, the role of collaborative and federated learning, and the significance of explainable AI and interpretability in detection techniques. We will also discuss opportunities to combine multiple detection techniques and fine-tune models over time. Finally, we propose the idea of using generative agents as source of synthetic training data. By highlighting these opportunities, we aim to provide a roadmap for researchers and practitioners to harness these advances and improve the effectiveness of social bot detection.

### *4.1. Leveraging transfer learning and unsupervised learning*

Transfer learning and unsupervised learning present valuable opportunities for enhancing the performance and efficiency of social bot detection techniques, addressing challenges such as data scarcity, expensive annotation processes, and adapting to the evolving behavior of bots.

Transfer learning is a technique that allows models to leverage knowledge gained from pre-trained models or related tasks, which can be fine-tuned for detecting social bots (Pan and Yang, 2010). For instance, pre-trained language models, such as BERT (Devlin, *et al.*, 2018) and GPT (Radford, *et al.*, 2019), have demonstrated the effectiveness of transfer learning in natural language processing tasks. By fine-tuning these models on smaller, task-specific datasets, researchers can exploit the rich semantic and syntactic information captured during pre-training to identify bot-generated content. This approach can help improve the accuracy of classifiers while reducing the need for large amounts of labeled training data. Furthermore, transfer learning can help mitigate the impact of domain shift, which occurs when the distribution of data used for training differs from that of target data (Torrey and Shavlik, 2010). By leveraging pre-trained models, researchers can adapt more effectively to the evolving behavior and strategies of social bots, as demonstrated by the work of Yang, *et al.* (2020) in detecting bots using BERT-based models.

Unsupervised learning techniques, which do not require labeled data, can be employed to identify patterns, anomalies, or clusters in the data that may be indicative of social bot activity (Chandola, *et al.*, 2009). Blending methods such as clustering, dimensionality reduction, and autoencoders can help capture the underlying structure of the data and reveal potential bot-like behavior without relying on labeled examples. For example, dimensionality reduction techniques, such as t-distributed stochastic neighbor embedding (t-SNE), can be used to visualize high-dimensional data, facilitating the identification of patterns and anomalies in user behavior (van der Maaten and Hinton, 2008). Combined with autoencoders — a type of neural network — they can be employed for unsupervised feature learning, capturing latent representations of data that may be proven useful for detecting bots (Haider, *et al.*, 2023).

In conclusion, leveraging transfer learning and unsupervised learning approaches can address the challenges associated with data scarcity, expensive annotation processes, and adapting to the evolving behavior of bots in social bot detection tasks. By exploiting the rich knowledge captured by pre-trained models and unsupervised techniques, researchers can develop more accurate and robust detection methods.

### *4.2. Multimodal approaches to detection*

Multimodal approaches, which combine different types of data, such as text, images, and network features, can provide a more comprehensive and robust understanding of user behavior, thereby improving the detection of social bots. By incorporating diverse sources of information, these techniques can capture complex and subtle patterns in the data that may be overlooked by unimodal approaches.

Recently, there has been a growing interest in using multimodal data for social bot detection. One promising direction is the integration of text and image analysis. For example, Wu, *et al.* (2019) proposed a multimodal approach to detecting fake news by jointly modeling textual and visual information. Their approach used a hierarchical attention network to capture the dependencies between textual and visual features, enabling the model to identify bots that disseminate misleading information through both text and visual content. Similarly, Besel, *et al.* (2018) proposed a model that combined textual and visual cues to detect social bots on Twitter, demonstrating the effectiveness of multimodal approaches in capturing complementary information.

Another direction for multimodal approaches is the incorporation of network features. Social bot detection can benefit from analyzing user interactions and network structures, which can reveal suspicious interaction patterns or network structures that may be indicative of coordinated bot activity (Pacheco, *et al.*, 2020; Pacheco, *et al.*, 2021). Sharma, *et al.*. (2021) demonstrated the value of temporal features by proposing a method that combined them with content-based features to coordinated influence campaigns on social media platforms.

Multimodal approaches can also be extended to other data types, such as audio or video, to further improve social bot detection capabilities. For example, audio or video analysis can be used to identify bots that generate or distribute deepfake content, which poses a significant threat to online platforms (Rössler, *et al.*, 2019). Combining these diverse sources of information can lead to more accurate and robust detection methods that are better equipped to handle the evolving nature of social bots.

In summary, multimodal approaches to social bot detection hold great promise for improving the efficacy of detection methods. By combining different types of data, such as text, images, and network features, researchers can capture complex and subtle patterns in the data that may be overlooked by unimodal approaches. This can lead to more accurate and robust social bot detection techniques, enabling more effective mitigation of the negative impact of social bots on online platforms.

### *4.3. Collaborative and federated learning for detection*

Collaborative and federated learning approaches can enable the joint learning and sharing of knowledge among multiple organizations or platforms while preserving data privacy. These methods allow the training of models on distributed data sources without requiring the data to be centralized, which can help overcome privacy concerns and data sharing limitations. By pooling the collective knowledge and resources of different organizations, these approaches can lead to more accurate and robust social bot detection techniques that can be deployed across various platforms and domains.

Federated learning is a distributed machine learning approach that enables multiple clients to collaboratively train a shared model while keeping their data locally (McMahan, *et al.*, 2016). This approach can be particularly useful for social bot detection, as it allows for the creation of a global model based on the collective knowledge of various organizations without compromising the privacy of user data. In a federated learning setup, each organization trains a local model using its data and shares only the model updates (*e.g.*, gradients) with a central server, which aggregates these updates to improve the global model. This process is iteratively repeated until convergence, resulting in a model that benefits from the diverse experiences of the participating organizations.

Several studies have explored the potential of federated learning for social bot detection. For example, Fung, *et al.* (2020) proposed a federated learning approach for detecting social bots in online social networks. They demonstrated that their approach could effectively detect bots while preserving data privacy, even in scenarios with non-IID (independently and identically distributed) data across clients. Similarly, Nguyen, *et al.* (2021) proposed a federated learning framework for social bot detection that leverages a hierarchical attention mechanism to capture the relationships between user features, content features, and social network features.

Collaborative learning is another approach to distributed learning that enables multiple organizations to jointly learn from their data while preserving privacy (Veale, *et al.*, 2018). In contrast to federated learning, which relies on a central server to aggregate model updates, collaborative learning techniques often rely on decentralized protocols, such as gossip learning or peer-to-peer networks, to exchange information and learn from one another. Collaborative learning can be particularly useful in situations where there is no trusted central server or when organizations want to maintain control over their data and learning processes.

Finally, collaborative and federated learning approaches hold significant potential for improving social bot detection across multiple organizations and platforms while preserving data privacy (Ezzeldin, *et al.*, 2021). By enabling the joint learning and sharing of knowledge without the need for centralizing data, these approaches can help overcome the limitations of traditional machine learning techniques and lead to more accurate and robust social bot detection methods.

### *4.4. Explainable AI and interpretability in detection techniques*

Explainable AI (XAI) and interpretability play a crucial role in improving the trustworthiness, transparency, and accountability of social bot detection techniques. By understanding the underlying decision-making process of AI models, stakeholders can gain insights into the factors that contribute to the detection of social bots, enabling more informed decisions and facilitating the improvement of detection methods. Furthermore, explainability can help address concerns about biases and ethical implications, which are particularly important in the context of online platforms and social media.

Interpretable models, such as decision trees, linear models, and rule-based systems, are inherently explainable, as their decision-making process can be easily understood by humans (Ribeiro, *et al.*, 2016). For instance, decision trees allow for the visualization of the decision-making process, with nodes representing feature splits and leaf nodes indicating class decisions (Quinlan, 1986). In the context of social bot detection, decision trees can help identify key features and decision rules that are indicative of bot-like behavior (Ferrara, *et al.*, 2016).

However, many state-of-the-art machine learning models, such as deep neural networks, are often considered black-box models due to their complex architectures and large number of parameters (Adadi and Berrada, 2018). To make these models more interpretable, several techniques have been proposed, including local interpretable model-agnostic explanations (LIME) and SHapley Additive exPlanations (SHAP).

LIME is an explanation technique that aims to provide local explanations for individual predictions of black-box models (Ribeiro, *et al.*, 2016). It works by approximating the local decision boundary of the model using interpretable models, such as linear models or decision trees. In the context of social bot detection, LIME can help identify the features that are most important for the classification of a specific user as a bot or human, thereby providing insights into the underlying decision-making process.

SHAP is another explanation technique that assigns importance values to each feature by calculating the contribution of each feature to the prediction for a given instance, based on cooperative game theory (Lundberg and Lee, 2017). By computing the Shapley values for each feature, SHAP can provide a consistent and fair allocation of the prediction's contribution across all features. In the context of social bot detection, SHAP can help uncover the most influential features that contribute to the detection of bots, facilitating an understanding and improvement of detection methods.

In conclusion, explainable AI and interpretability are essential for improving the trustworthiness, transparency, and accountability of social bot detection techniques. By providing insights into the decision-making process of AI models, stakeholders can make more informed decisions, improve detection methods, and address concerns about biases and ethical implications. Future research should focus on developing more explainable and interpretable models for social bot detection and exploring the interplay between explainability and model performance.

### *4.5. Combining multiple detection techniques*

Effectively detecting social bots necessitates a comprehensive approach that considers the diverse behaviors and strategies employed by these malicious entities. Integrating multiple detection techniques can lead to more accurate and robust results, addressing a broader range of bot features and behaviors. In this section, we explore various detection techniques and their integration, while referencing relevant research papers.

One approach that has been widely adopted in social bot detection is the combination of machine learning and natural language processing techniques. By incorporating both supervised and unsupervised learning algorithms, as well as natural language processing methods, researchers can analyze content generated by social bots and identify patterns and features that differentiate them from human users (Ferrara, *et al.*, 2016). For instance, Yang, *et al.* (2019) proposed a deep learning-based method that combined convolutional neural networks (CNNs) and recurrent neural networks (RNNs) to analyze content and account features for bot detection, achieving high accuracy and recall.

Network analysis is another essential technique for detecting social bots, as it focuses on analyzing the social network structures and interaction patterns of bot accounts. By investigating these patterns, researchers can identify anomalous behavior and uncover coordinated activities, such as astroturfing campaigns and disinformation efforts (Ratkiewicz, *et al.*, 2011). Temporal analysis has also proven to be a valuable tool for uncovering social bot behavior. By investigating the temporal dynamics of bot activities, such as posting frequency and the timing of interactions, researchers can identify unusual patterns that may indicate automated behavior (Chavoshi, *et al.*, 2016). In their research, Chavoshi, *et al.* (2016) applied temporal analysis to detect social bots on Twitter by examining the warped correlation between user activity and content generation. Their approach successfully identified bots with high precision and recall, demonstrating the effectiveness of temporal analysis in bot detection.

Cross-platform analysis is another promising technique for social bot detection, as it involves examining the behavior of suspected bot accounts across multiple social media platforms. By detecting coordinated activities and improving the generalizability of detection techniques, researchers can develop more effective and adaptable bot detection methods (Zhou and Zafarani, 2018). Shu, *et al.* (2018) introduced a cross-platform framework called FakeNewsNet to study the diffusion of fake news and the role of social bots in its spread. By collecting a large dataset of news articles labeled as fake or real, and analyzing the activity of social bots across platforms, they were able to identify patterns of fake news dissemination and the characteristics of bots involved in spreading it.

In summary, combining multiple detection techniques, including machine learning and natural language processing, network analysis, temporal analysis, and cross-platform analysis, can lead to more accurate and robust social bot detection results.

### *4.6. Constantly updating and fine-tuning models*

To effectively counter the ever-evolving landscape of social bots, it is vital to maintain a proactive approach in the development and deployment of detection models. This requires the continuous update and fine-tuning of models in response to the changing tactics and strategies employed by bot operators. In this section, we delve into the various aspects of model maintenance and improvement, with reference to relevant research papers.

Ensuring that the training data used for machine learning algorithms is up-to-date and representative of the current bot landscape is critical for maintaining the effectiveness of detection models. This may involve the collection of new data, re-labeling of existing data, or the incorporation of external data sources, such as labeled datasets from other researchers or organizations (Ruchansky, *et al.*, 2017). For example, Cresci, *et al.* (2017) demonstrated the importance of using diverse and recent datasets for training and evaluating social bot detection models. They conducted a systematic comparison of multiple datasets and found that the performance of models trained on older datasets was significantly lower than those trained on more recent and diverse data.

Adapting to platform changes is another crucial aspect of maintaining detection models. Social media platforms are constantly evolving, with updates to APIs or user interfaces that may impact the availability of certain features or introduce new patterns of bot behavior. Monitoring these changes and adjusting detection techniques accordingly can help ensure the continued effectiveness of detection models (Gorwa, *et al.*, 2020). As an example, Echeverria and Zhou (2017) examined the impact of Twitter API rate limits on social bot detection and found that the imposed restrictions reduced the accuracy of certain detection features, highlighting the need to adapt models to accommodate platform changes.

Fine-tuning model parameters is an essential part of maintaining the effectiveness of detection models. Regularly evaluating the performance of these models and adjusting their parameters to optimize accuracy, recall, and other relevant metrics can lead to improved results. This may involve experimentation with different feature sets, algorithms, or parameter values, and conducting cross-validation to assess the robustness of the models. Kudugunta and Ferrara (2018) introduced a deep learning model based on contextual LSTM networks to detect social bots on Twitter. They fine-tuned various hyperparameters of the model and assessed its performance through cross-validation, demonstrating the importance of parameter optimization for achieving the best results.

Incorporating feedback loops is another vital practice for maintaining detection models. Establishing feedback mechanisms to collect user input on the accuracy of bot detection results can help refine models and improve their performance. This may involve crowdsourcing, expert validation, or other forms of user engagement to gather insights and evaluate the effectiveness of detection techniques. For instance, Yang, *et al.* (2022) developed a system called Botometer, which combined machine learning-based bot detection with user feedback to improve its performance. The system allowed users to report false positives and false negatives, and the input was used to fine-tune the underlying detection models.

In conclusion, constantly updating and fine-tuning social bot detection models is crucial for staying ahead of the evolving landscape of social bots. By adopting practices such as updating training data, adapting to platform changes, fine-tuning model parameters, and incorporating feedback loops, researchers and practitioners can develop more effective and resilient detection techniques that better protect online information ecosystems.

### *4.7. Generative agents as source of synthetic data*

The field of social bot detection can greatly benefit from leveraging synthetic data, especially data generated using believable proxies of human behavior. We discuss the potential of generative agents—computational software agents that simulate believable human behavior—to enhance social bot detection techniques by providing a rich source of synthetic data based on large language models.

Generative agents, which have been explored in recent literature, are designed to exhibit human-like behavior by storing a complete record of the agent's experiences in natural language, synthesizing memories over time into higher-level reflections, and dynamically retrieving them to plan behavior (Park, *et al.*, 2023). These agents can produce believable individual and emergent social behaviors (see Figure 1), making them a valuable source of synthetic data for training and evaluating social bot detection models.

*Figure 1. Generative agents and their simulated interactions (from Park et al., 2023). This type of simulations could prove invaluable as a framework for the scalable generation of realistic synthetic data that can be used to train bot detection models.*

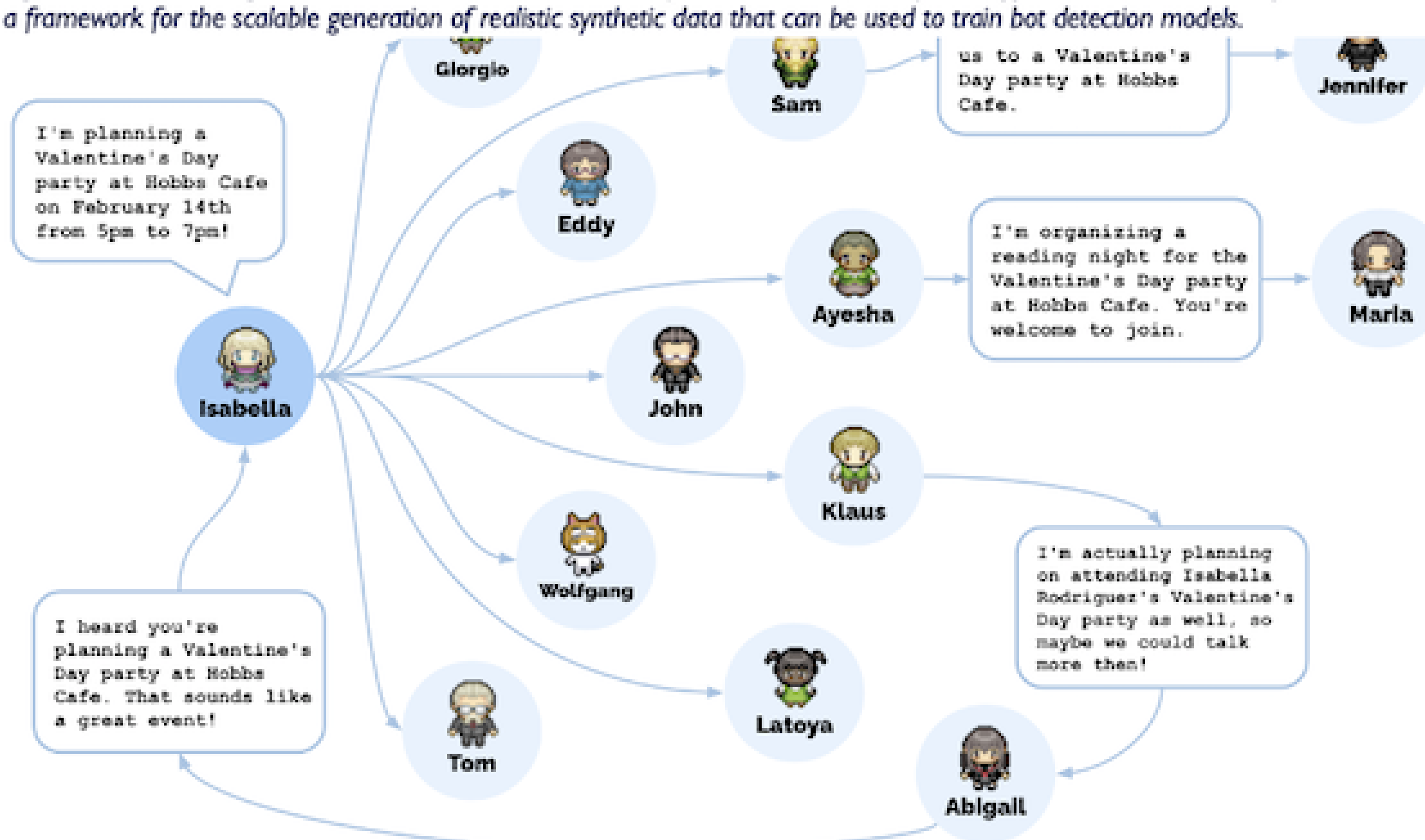

By fusing large language models with computational, interactive agents, researchers can generate vast amounts of realistic and diverse data that can be used to train more robust and accurate detection models. The synthetic data created by generative agents can be utilized to simulate various scenarios and communication patterns, allowing researchers to develop detection models that can generalize well across different contexts.

Moreover, the use of synthetic data created by generative agents can help address some of the challenges faced in social bot detection, such as the scarcity of labeled data and the constantly evolving nature of bot-generated content. Since generative agents can be controlled and manipulated, researchers can generate labeled data at scale, effectively reducing the reliance on manual annotation and data collection processes.

Furthermore, generative agents can be updated and fine-tuned over time to mimic the latest trends and patterns observed in bot-generated content, ensuring that the synthetic data remains relevant and useful for training and evaluating detection models. This adaptability can be crucial in the ongoing battle against increasingly sophisticated social bots.

In conclusion, the use of generative agents and synthetic data offers promising opportunities for enhancing social bot detection techniques. By leveraging large language models and computational agents to generate realistic and diverse data, researchers can develop more robust and accurate detection models that are better equipped to identify and mitigate the impact of malicious social bots.

## 5. Case studies: Social bot detection in real-world applications

In this section, we present a series of case studies illustrating the real-world applications of social bot detection techniques across various domains. These examples demonstrate the significance of social bot detection in addressing critical issues such as election interference and political manipulation, disinformation campaigns and fake news, and financial scams and cryptocurrency manipulation. By examining these case studies, we aim to shed light on the practical implications of social bot detection and the importance of continued research and development in this field.

### *5.1. Election interference and political manipulation*

Social bots have been widely used to manipulate public opinion and interfere with elections, as they can easily disseminate political propaganda, false information, and polarizing content. The 2016 U.S. presidential election is a well-known example, where social bots played a significant role in amplifying and spreading politically biased content, affecting the dynamics of the election (Ferrara, 2015; Bessi and Ferrara, 2016; Badawy, *et al.*, 2018; Chang, *et al.*, 2021). Researchers have developed various social bot detection techniques specifically targeting political bots, aiming to minimize their influence on public discourse and democratic processes (Gorwa, *et al.*, 2020; Howard and Kollanyi, 2016).

Bessi and Ferrara (2016) conducted an in-depth analysis of the 2016 U.S. presidential election, revealing that social bots were responsible for generating and disseminating a significant portion of the election-related content on social media platforms, particularly Twitter. They observed that bots generated approximately one-fifth of the entire conversation around the election, with a strong bias towards specific political topics and candidates. By employing machine learning algorithms to identify bot accounts based on their behavioral patterns, Bessi and Ferrara were able to uncover the extensive presence of bots and their potential influence on the election. These findings were later connected to a state-sponsored operation led by Russia to interfere with the U.S. election (Badawy, *et al.*, 2018; Addawood, *et al.*, 2019; Luceri, *et al.*, 2019).

Another study by Howard and Kollanyi (2016) investigated the role of social bots during the 2016 Brexit referendum in the United Kingdom. They found that bots contributed to a significant portion of the Twitter traffic related to the referendum, with pro-Leave bots being more active than pro-Remain bots. The researchers employed supervised machine learning techniques to classify bot accounts, using features such as tweet frequency, retweet ratio, and content similarity. Their findings highlighted the potential impact of social bots on shaping public opinion during critical political events.

Stella, *et al.* (2018) proposed a method to detect social bots in the context of political discussions on Twitter, focusing on the 2017 Catalan referendum for independence. They used a combination of unsupervised learning techniques, such as clustering and dimensionality reduction, to identify groups of similar users and detect bots based on their behavioral patterns. By analyzing the content produced by these bots, Stella, *et al.* were able to uncover coordinated disinformation campaigns and provide insights into the strategies employed by bot operators during the election.

In conclusion, social bots have played a significant role in manipulating public opinion and interfering with elections in various countries. Research on detecting and understanding the behavior of political bots has provided valuable insights into their strategies and potential impact on democratic processes. By developing effective detection techniques and countermeasures, it is possible to mitigate the influence of social bots on public discourse and protect the integrity of electoral processes.

### *5.2. Detection of disinformation campaigns by social bots*

The spread of disinformation and fake news through social media platforms has become a significant concern, as it can lead to misinformation, increased polarization, and decreased trust in media and institutions. Social bots can exacerbate this issue by rapidly disseminating false information and creating the illusion of widespread support for a particular narrative (Ferrara, 2017; Shao, *et al.*, 2018). Detecting social bots involved in disinformation campaigns is crucial for mitigating the spread of fake news and maintaining the integrity of online information.

We analyzed the role of social bots in shaping public opinion during the French presidential elections (Ferrara, 2017). Our investigation on the spread of disinformation, particularly on Twitter, uncovered the significant presence of automated accounts disseminating politically biased content. The study highlighted the extensive use of social bots in promoting distorted narratives, manipulating public discourse, and polarizing the online conversation.

Shao, *et al.* (2018) conducted a study on the role of social bots in spreading low-credibility content on Twitter during and after the 2016 U.S. presidential election. They discovered that social bots played a significant role in amplifying low-credibility content, with some bots even operating in a coordinated fashion. By employing machine learning techniques, Shao, *et al.* were able to detect and characterize the behavior of bots involved in spreading disinformation, providing valuable insights into their strategies and potential impact on public discourse.

In a study by Ruchansky, *et al.* (2017), the authors proposed a hybrid deep-learning model called CSI (Capture, Score, and Integrate) to detect fake news and the social bots responsible for its spread. By leveraging features from both content and social network structures, their model achieved high accuracy in detecting both fake news and the bots disseminating it, demonstrating the potential of AI-based techniques in addressing disinformation challenges.

In their study, Vosoughi, *et al.* (2018) explored the spread of true and false news online and found that false information spread more rapidly and broadly than true information, partially due to the activity of social bots. They used a variety of machine learning techniques to model the spread of news and identify features that differentiated true from false information. Their findings underscored the need for effective social bot detection and countermeasures to limit the spread of disinformation.

Tackling the issue of disinformation and fake news requires a multifaceted approach, including the development of social bot detection techniques that can identify and mitigate the activity of bots involved in spreading false information. By leveraging advanced machine learning and natural language processing methods, researchers can better understand the strategies and tactics employed by disinformation campaigns and design interventions to protect the integrity of online information sources.

### *5.3. Financial scams and cryptocurrency manipulation*

The proliferation of social bots on social media platforms has led to an increase in financial scams and cryptocurrency manipulation. In this section, we delve into the technical details of how social bots contribute to these issues, discussing their strategies and potential detection methods. We also reference relevant published papers to provide an in-depth understanding of this growing concern.

Social bots can be programmed to execute various financial scams, such as pump-and-dump schemes, phishing, and spreading false information to manipulate stock prices or cryptocurrency values. These bots often mimic human behavior and can spread misinformation at an alarming rate, making them particularly effective at deceiving unsuspecting users.

In the realm of cryptocurrency, social bots have been found to engage in market manipulation by influencing public sentiment. By generating a large volume of fake messages promoting certain cryptocurrencies or spreading false information about market trends, these bots can create artificial demand, driving up the value of targeted cryptocurrencies. Once the value reaches a predetermined threshold, the bots or their controllers can sell off their holdings, causing a rapid decline in a cryptocurrency’s value and leaving other investors at a loss (Vasek and Moore, 2015).

Detecting social bots involved in financial scams and cryptocurrency manipulation can be challenging due to their ability to adapt their behavior and evade traditional detection methods. However, several studies have proposed innovative techniques to tackle this issue. For example, Nizzoli and collaborators (2020) developed a data-driven framework to identify pump-and-dump schemes in cryptocurrency markets. By collecting and analyzing millions of messages from platforms like Twitter, Telegram, and Discord, their model was able to unveil two mechanisms (pump-and-dump and Ponzi schemes) revealing deceptive activity associated with suspicious bot accounts involved in cryptocurrency frauds.

Another study by Nghiem, *et al.* (2018) proposed a method for detecting cryptocurrency-related social bots on Twitter. The authors employed a combination of network analysis, content analysis, and machine learning techniques to identify bot accounts that were promoting or spreading false information about cryptocurrencies. This multi-faceted approach proved effective at detecting bots involved in cryptocurrency manipulation, even in the presence of sophisticated evasion tactics.

To mitigate the impact of social bots in financial scams and cryptocurrency manipulation, it is crucial for researchers to develop novel detection techniques that can adapt to the ever-evolving strategies employed by these bots. Potential research directions include the incorporation of deep learning and reinforcement learning algorithms, as well as the development of models that can analyze multimodal data sources, such as text, images, and videos (Ferrara, 2022).

## 6. Conclusions

Social bot detection has emerged as an increasingly important area of research, as malicious bots continue to impact online information ecosystems and pose significant issues in various domains. In this concluding section, we provide a recap of the challenges and opportunities in social bot detection, discuss the future of detection in the age of ChatGPT, and present final thoughts and potential research directions.

### *6.1. Recap of challenges and opportunities in social bot detection*

In this paper, we have explored various challenges and opportunities in the realm of social bot detection, as well as the impact of advanced AI-generated chatbots, such as ChatGPT, on the field. Throughout the discussion, we have drawn upon numerous relevant published papers to provide a comprehensive and technically detailed perspective on the subject.

Social bot detection faces several challenges. One significant challenge is the increasing sophistication of AI-generated content, which makes it harder to differentiate between human users and social bots. Moreover, adversarial attacks and evasion tactics further complicate the detection process, as bots become more adept at evading traditional detection methods. The need for scalable and real-time detection solutions is another challenge, given the vast scale and dynamic nature of social media platforms. Lastly, the ethical considerations and privacy concerns associated with social bot detection pose unique challenges that need to be addressed to maintain public trust and adhere to legal and ethical frameworks.

Despite these challenges, numerous opportunities exist for improving social bot detection techniques. Transfer learning and unsupervised learning offer promising approaches to leverage existing knowledge and discover underlying patterns in bot behavior. Multimodal approaches to detection can capitalize on the integration of diverse data sources and features, enabling more robust and accurate detection. Collaborative and federated learning for detection can help facilitate cooperation between different organizations and platforms, improving the overall effectiveness of detection methods. Finally, the incorporation of explainable AI and interpretability in detection techniques can help to build trust and provide valuable insights into the underlying decision-making processes of these models.

### *6.2. The future of social bot detection in the age of ChatGPT*

As we progress into the age of ChatGPT and other advanced language models, the future of social bot detection will need to adapt to the ever-increasing sophistication of AI-generated content. Researchers and practitioners must remain vigilant in the face of rapid advancements in AI-generated chatbots, which continually blur the distinction between human- and machine-generated content (Radford, *et al.*, 2019). This section discusses the future of social bot detection, focusing on the implications of advanced AI-generated chatbots and the potential pathways for detection methods to evolve in response.

To effectively counteract the improved capabilities of AI-generated chatbots, researchers must explore new methodologies and approaches that complement existing techniques. One potential direction is the development of more sophisticated machine learning models that can account for the subtleties and nuances in the content produced by advanced chatbots like ChatGPT. These models should be able to discern not only the linguistic features but also the contextual and behavioral aspects associated with bot-generated content.

Transfer learning and adversarial training (Goodfellow, *et al.*, 2014) present promising avenues for improving the robustness and generalizability of social bot detection models. Transfer learning can help in leveraging pre-trained models and knowledge from other domains to enhance bot detection capabilities. Adversarial training, on the other hand, can introduce adversarial examples during the model training process, improving the model’s resilience to evasion techniques employed by bots.

Incorporating contextual and multimodal information into detection methods is another potential pathway to enhance the capabilities of bot detection techniques. By analyzing not only textual content but also images, videos, and other media types, as well as considering user behavior and network structures, researchers can develop more comprehensive and robust detection models that account for a wider range of bot characteristics and behaviors.

As AI-generated chatbots continue to evolve, it is crucial to consider the ethical implications of social bot detection. The development of guidelines and policies that address the ethical, legal, and societal dimensions of the issue will be necessary to ensure that detection methods respect user privacy and adhere to legal frameworks.

In conclusion, the future of social bot detection in the age of ChatGPT and other advanced language models will require innovative techniques that can adapt to the rapidly changing landscape of AI-generated content. By exploring new methodologies and approaches, fostering interdisciplinary collaboration, and addressing the ethical implications of detection, researchers can contribute to a more secure and trustworthy online environment.

### *6.3. Final remarks and potential research directions*

Throughout this paper, we have explored the challenges and opportunities in social bot detection, focusing on the implications of advanced AI-generated chatbots such as ChatGPT. In this final section, we outline potential research directions that can help advance the field of social bot detection and contribute to the development of more effective and reliable detection techniques.

- ***Human-AI collaboration***: As AI-generated chatbots become more sophisticated, it is essential to consider the potential of human-AI collaboration in social bot detection. Subject matter expertise can help guide and refine detection models, while AI can provide the computational capabilities necessary for analyzing vast amounts of data. Investigating new paradigms for human-AI collaboration, such as active learning and mixed-initiative systems, could yield valuable insights and improvements in social bot detection performance.

- ***Cross-platform detection***: Most research on social bot detection has focused on specific platforms, such as Twitter or Facebook. However, as new social media platforms continue to emerge, it is crucial to develop detection methods that can generalize across platforms. This would involve the creation of cross-platform datasets and the development of models that can account for platform-specific features and user behaviors.

- ***Temporal dynamics of bots***: Many existing detection methods are static, focusing on a single snapshot of bot behavior. However, bots may change their behavior over time, either as part of their strategy or in response to detection efforts (Luceri, *et al.*, 2019; Luceri, *et al.*, 2020). Future research should explore the temporal dynamics of social bots, incorporating time-series analysis and other techniques for tracking and detecting evolving bot behavior (Pozzana and Ferrara, 2020).

- ***Counterfactual reasoning***: As AI-generated chatbots improve, they may become capable of generating content that closely mimics human behavior. In such cases, it may be helpful to explore counterfactual reasoning as a means of detecting bots. By considering what a bot would do in each situation compared to a human user, researchers may be able to develop new and more effective detection techniques.

- ***Low-resource settings***: Investigating and developing novel techniques for social bot detection in low-resource languages, leveraging cross-lingual transfer learning, unsupervised learning, and data augmentation methods to improve detection capabilities can support underrepresented communities in combating malicious social bots.

- ***Multimodal detection***: Exploring the integration of diverse data modalities, such as text, images, audio, and video, in social bot detection frameworks to enhance the robustness and accuracy of detection algorithms, can address the limitations of single-modality approaches in identifying complex and sophisticated social bots.

- ***Privacy-preserving detection techniques***: With increasing concerns about user privacy and data protection, it is essential to develop detection techniques that can operate within privacy-preserving frameworks. Techniques such as differential privacy and federated learning could be employed to ensure that detection methods do not compromise user privacy while still effectively identifying malicious bots.

In conclusion, the field of social bot detection offers numerous opportunities for researchers to explore and develop innovative techniques to counteract the growing sophistication of AI-generated chatbots. By addressing these potential research directions and fostering collaboration between researchers, practitioners, and policy-makers, we can work towards safer and trustworthy online environments for all users.

## About the author

Emilio Ferrara is Professor at the University of Southern California, Research Team Leader at the USC Information Sciences Institute, and Principal Investigator at the USC/ISI Machine Intelligence and Data Science (MINDS).
E-mail: emiliofe [at] usc [dot] edu

## Acknowledgements

The author is indebted with all current and past members of his lab at USC, core researchers, and visiting students at ISI, collaborators at Indiana University, and all coauthors of bot-related work: without them, most advancements in this research field would have been impossible.

---

---